\documentclass[aps,prl,amsmath,amssymb,bibnotes,reprint,longbibliography, superscriptaddress]{revtex4-2}
\usepackage{t1enc}
\usepackage[utf8]{inputenc}
\usepackage{amsmath}

\usepackage{graphicx}
\usepackage{siunitx}
\usepackage{hyperref}

\begin{document}

\title{Triplet-blockaded Josephson supercurrent in double quantum dots}

\author{Dani\"{e}l Bouman}
\author{Ruben J. J. van Gulik}
\affiliation{QuTech and Kavli Institute of Nanoscience, Delft University of Technology, NL-2600 GA Delft, The Netherlands}
\author{Gorm Steffensen}
\affiliation{Center for Quantum Devices, Niels Bohr Institute, University of Copenhagen, DK-2100 Copenhagen, Denmark}
\author{D\'{a}vid Pataki}
\affiliation{Department of Theoretical Physics and MTA-BME Exotic Quantum Phases Research Group, Budapest University of Technology and Economics, H-1111 Budapest, Hungary}
\author{P\'{e}ter Boross}
\affiliation{Institute for Solid State Physics and Optics, Wigner Research Centre for Physics, P.O. Box 49, H-1525 Budapest, Hungary}
\author{Peter Krogstrup}
\author{Jesper Nyg\r{a}rd}
\author{Jens Paaske}
\affiliation{Center for Quantum Devices, Niels Bohr Institute, University of Copenhagen, DK-2100 Copenhagen, Denmark}
\author{Andr\'{a}s P\'{a}lyi}
\affiliation{Department of Theoretical Physics and MTA-BME Exotic Quantum Phases Research Group, Budapest University of Technology and Economics, H-1111 Budapest, Hungary}
\author{Attila Geresdi}
\email[Corresponding author. E-mail address: ]{geresdi@chalmers.se}
\affiliation{QuTech and Kavli Institute of Nanoscience, Delft University of Technology, NL-2600 GA Delft, The Netherlands}
\affiliation{Quantum Device Physics Laboratory, Department of Microtechnology and Nanoscience, Chalmers University of Technology, SE-41296 Gothenburg, Sweden}

\begin{abstract}
Serial double quantum dots created in semiconductor nanostructures provide a
versatile platform for investigating two-electron spin quantum states, which can
be tuned by electrostatic gating and an external magnetic field. In this work, we
directly measure the supercurrent reversal between adjacent charge states of an
InAs nanowire double quantum dot with superconducting leads, in good agreement
with theoretical models. In the even charge parity sector, we observe a
supercurrent blockade with increasing magnetic field, corresponding to the spin
singlet to triplet transition. Our results demonstrate a direct spin to
supercurrent conversion, the superconducting equivalent of the Pauli spin
blockade. This effect can be exploited in hybrid quantum architectures coupling the
quantum states of spin systems and superconducting circuits.
\end{abstract}

\maketitle

Semiconductor quantum dots, where the orbital and spin states of single
localized electrons can be controlled \cite{RevModPhys.79.1217}, are one of the
leading platforms for quantum information processing \cite{loss1998quantum}.
Specifically, double quantum dots (DQDs) connected in a series
\cite{van2002electron} became the preferred physical implementation of spin
\cite{Nowack1430}, and spin-orbit quantum bits \cite{NadjPerge2010} in
low-dimensional semiconductor nanodevices, such as heterostructures hosting a
two-dimensional electron gas or semiconductor nanowires. In these devices, the
readout of the spin quantum state relies on spin-dependent single electron
tunneling processes, which then enable charge
readout via direct electronic transport \cite{RevModPhys.79.1217}, charge
sensing techniques \cite{PhysRevLett.70.1311}, or dipole coupling to a microwave
resonator \cite{PhysRevLett.108.046807, Petersson2012}.

In a superconducting nanodevice, the dissipationless supercurrent $I_\mathrm{S}$ at zero
voltage bias is driven by the quantum mechanical phase difference $\varphi$ up to
a maximum amplitude, $I_\mathrm{C}$, the critical current \cite{josephson}. In the lowest
order of tunneling, the supercurrent-phase relationship (CPR)
\cite{RevModPhys.76.411} is sinusoidal, $I_\mathrm{S}(\varphi)=I_\mathrm{C} \sin(\varphi)$, which
describes the coherent transfer of single Cooper pairs through the weak
link. When the weak link is a non-magnetic tunnel barrier, a zero phase
difference is energetically favorable in the absence of supercurrent, which is
described by a positive critical current, $I_\mathrm{C} > 0$. In contrast, a negative
coupling yields a supercurrent reversal, $I_\mathrm{C} < 0$, often denoted a $\pi$
junction due the $\pi$ phase shift in the CPR. This negative coupling has been
observed in ferromagnetic weak links \cite{PhysRevLett.86.2427, PhysRevLett.89.137007},
out-of-equilibrium electron systems \cite{Baselmans1999} and semiconductor
quantum dot junctions \cite{Dam2006, PhysRevB.93.195437}.

The dependence of the critical current on the spin-state and charge state of a DQD
has also been addressed theoretically \cite{PhysRevB.62.13569,
PhysRevB.66.085306, doi:10.1080/00018732.2011.624266, PhysRevB.84.134512, droste2012josephson,
 brunetti2013anomalous, pokorn2019second}, and the recent progress in materials science of
superconductor-semiconductor hybrid nanostructures \cite{krogstrup2015epitaxy}
enabled measurements of the amplitude of the critical current as well
\cite{szombati2016josephson, PhysRevLett.121.257701}, in correlation with the
charge states of the DQD.

In this Rapid Communication, we report on direct measurements of the CPR through a DQD weak
link formed by an electrostatically gated InAs nanowire. By employing a
phase-sensitive measurement scheme, where the DQD is embedded in a
superconducting quantum interference device (SQUID), we characterize the full
CPR of the DQD, enabling a signful measurement of $I_\mathrm{C}$. The direct observation
of the supercurrent reversal in the total charge number boundaries allowed us to
identify the even and odd occupied states. Finally, the magnetic-field
dependence of the supercurrent amplitude in the even occupied state reveals the
presence of a supercurrent blockade in the spin-triplet ground state, in
agreement with numerical calculations.

\begin{figure*}
\includegraphics[width=\linewidth]{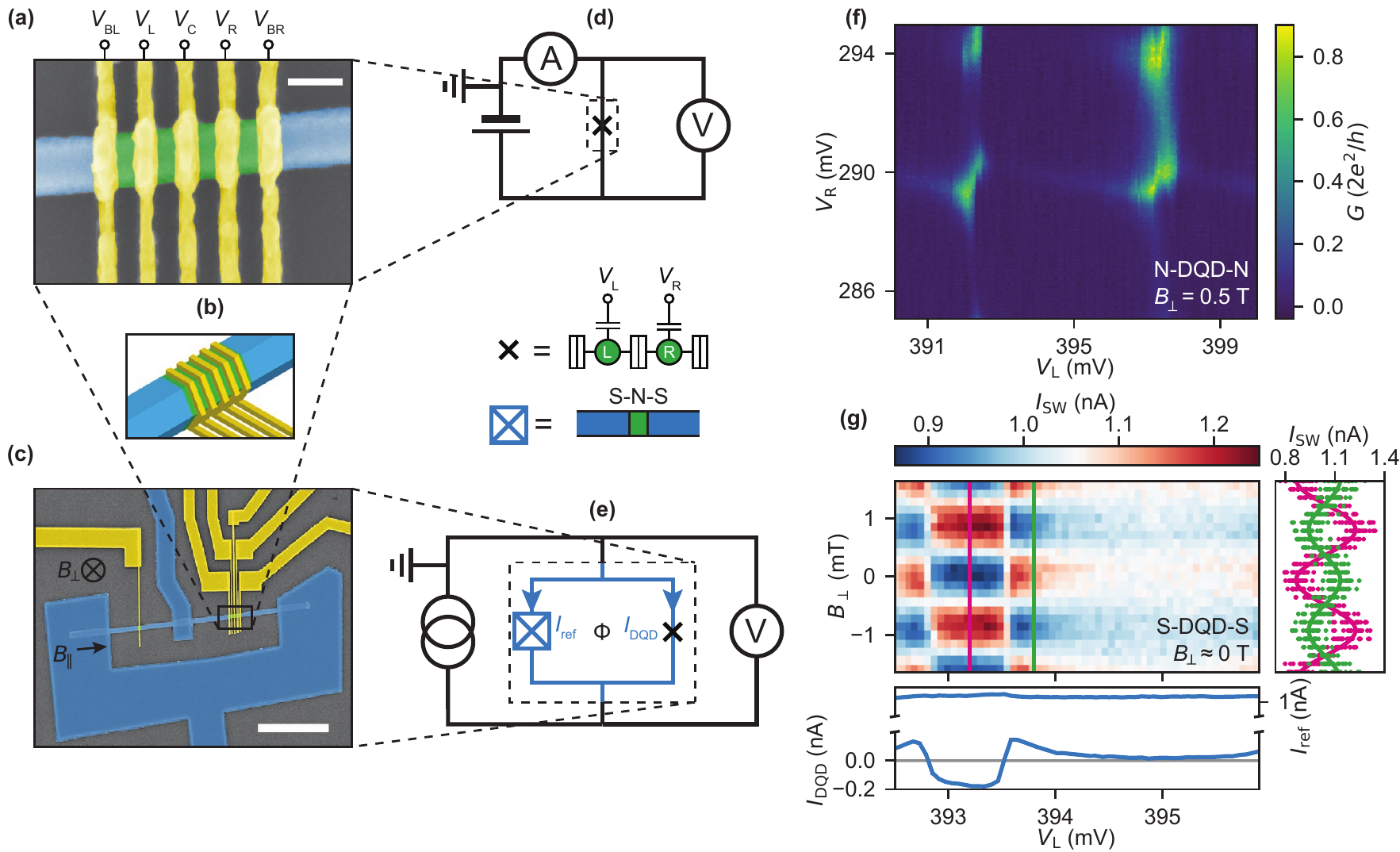}
\caption{Device layout and characterization. (a)
Color-enhanced electron micrograph of the nanowire DQD junction with five
wrap-around gates (yellow) which provide the confining potential. The $V_
\mathrm{BL}, V_\mathrm{C}$, and
$V_\mathrm{BR}$ gate voltages define the barriers, while $V_\mathrm{L}$ and $V_\mathrm{R}$ control the number of
electrons on the dots.
The aluminum shell (blue) is selectively etched away in the weak link section
(green denotes bare InAs). The scale bar denotes $100\,$nm.
(b) Perspective drawing of the DQD junction highlighting the conformal
gates. (c) Color-enhanced electron micrograph of the DC
SQUID made of sputtered NbTiN film (in blue) with the reference junction in
the left arm and the DQD junction in the right arm. The scale bar denotes
$2\,\mu$m. (d) The circuit diagram for the normal-state
characterization with the reference arm depleted.
(e) The measurement scheme of the switching current measurements in the
SQUID geometry.
(f) Charge stability diagram of the DQD in the normal state at a large magnetic
field $B_\perp = 0.5\,$T. (g) Switching current color map through three charge
states of the DQD and the flux $\Phi$ induced by a small $B_\perp$. Each pixel
is an average of $18$ measurements.
The side panel shows all switching current data taken along the magenta and
green line, respectively. The solid lines denote the sinusoidal fit yielding the
signful oscillation amplitude $I_\mathrm{DQD}$ and offset $I_\mathrm{ref}$ (see
text). The standard deviation of the phase is $6\times10^{-3}\pi$ and
$1.2\times10^{-2}\pi$ for the magenta and green lines, respectively. The bottom
panel displays the fitted $I_\mathrm{ref}$ and $I_\mathrm{DQD}$. The DQD was
tuned along the total energy axis [see the solid black line in Fig.~2(a)] and we
display the corresponding $V_\mathrm{L}$ range on the horizontal axis.} \label{fig1}
\end{figure*}

We built our device (Fig.~1) from an approximately $7\,\mu$m long InAs nanowire
grown by molecular beam epitaxy, and \emph{in-situ} partially covered by a
$6\,$nm thick epitaxial aluminum shell with a typical superconducting gap of
$\Delta \approx 200\,\mu$eV \cite{krogstrup2015epitaxy, van2017microwave}.
We formed two segments with the aluminum layer selectively removed where the DQD
and the reference arm would be defined. Next, we created the SQUID loop from a
sputtered NbTiN superconducting film, and covered the device with a
\SI{10}{\nano\meter} thick AlO$_x$ dielectric by conformal atomic layer
deposition. Finally, $40\,$nm wide and $50\,$nm thick Ti/Au gates [in yellow in
Fig.~1(a)] were evaporated under three angles to ensure a conformal coverage
around the wire [schematically shown in Fig.~1(b)]. Five gates defined the DQD (on
the right) and a single gate controlled the reference arm [on the left in
Fig.~1(c)]. Details on the device fabrication are shown in the Supplementary
Information \cite{supplement}. All of our measurements were performed in a dilution refrigerator
with a base temperature of approximately $30\,$mK.

We first characterize the DQD with the leads driven to the normal state by a
large magnetic field, $B_\perp=0.5\,$T. We measure the differential conductance
$\mathrm{d}I/\mathrm{d}V$ of the DQD with the reference arm fully depleted
[Fig.~1(d)]. We control the coupling to the leads with the gate voltages
$V_\mathrm{BL}$ and $V_\mathrm{BR}$, and the interdot coupling is tuned by
$V_\mathrm{C}$ [Fig.~1(a)]. A characteristic honeycomb diagram is plotted in
Fig.~1(f), where the charge occupancy of the dots $(n_\mathrm{L}, n_\mathrm{R})$
is set by the voltages applied on the two plunger gates, $V_\mathrm{L}$ and
$V_\mathrm{R}$.

We perform the CPR measurements with the leads being superconducting and with
the reference arm of the SQUID opened with its electrostatic gate so that it
exhibits a higher critical current than the DQD arm. Due to this asymmetry, the
phase drop over the DQD junction is determined by the magnetic flux $\Phi$
through the SQUID loop area [Fig.~1(e)] \cite{Dam2006, PhysRevLett.99.127005},
which is proportional to the applied magnetic field $B_\perp$. We measure the
switching current $I_\mathrm{SW}$ of the SQUID by ramping a current bias in a
sawtooth waveform and recording the bias current value when the junction
switches to the resistive state marked by a threshold voltage drop of the order
of $10\,\mu$V. We show a typical data set in Fig.~1(g), where each pixel in the
main panel is an average of $18$ measurements. The right sidepanel shows the raw
datapoints at two plunger gate settings denoted by the magenta and green lines
in the main panel, as well as the fitted sinusoidal curves in the following
functional form,
\begin{equation}
I_\mathrm{SW}=I_\mathrm{ref}+I_\mathrm{DQD}\sin{\varphi},
\end{equation}
where $\varphi=2\pi (B_\perp-B_\mathrm{o})/B_\mathrm{p}$, with $B_\mathrm{p}
\approx 1.7\,$mT being the magnetic field periodicity corresponding to a flux
change equal to the superconducting flux quantum $\Phi_0=h/2e$ and
$B_\mathrm{o}$ being the offset perpendicular magnetic field. The switching
current values $I_\mathrm{ref}$ and $I_\mathrm{DQD}$ represent the reference arm
and the DQD junction contributions, respectively. We show these fitted values as
a function of the gate voltage $V_\mathrm{L}$ in the lower subpanel of Fig.~1(g),
which displays the sign change of $I_\mathrm{DQD}$ at the charge state
boundaries. We note that the change in the environmental impedance
\cite{ivanchenko1969josephson} causes a slight modulation of $I_\mathrm{ref}$ as
well, despite the lack of any capacitive coupling between the two weak links.
However, in our measurements $I_\mathrm{ref}>5|I_\mathrm{DQD}|$ is always
fulfilled, enabling a reliable observation of the supercurrent reversal in the
DQD.

\begin{figure}
\includegraphics[width=\linewidth]{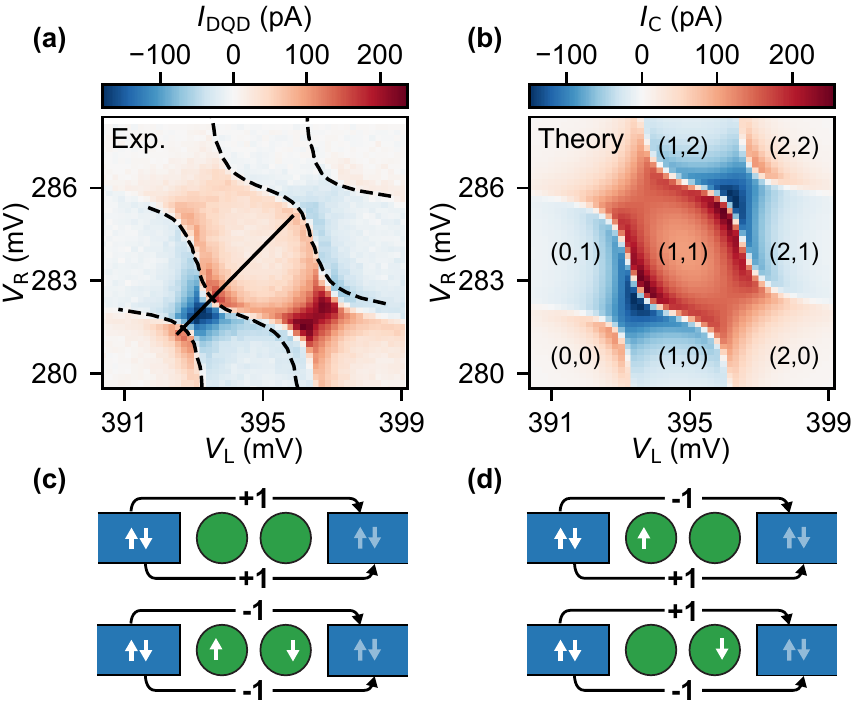}
\caption{The supercurrent charge-stability diagram at zero magnetic field. (a)
Colormap of the measured $I_\mathrm{DQD}$ as a function of the plunger gate
voltages $V_\mathrm{L}$ and $V_\mathrm{R}$ revealing a supercurrent sign reversal between the
adjacent total charge sectors. The dashed lines denote the numerically
calculated charge boundaries (see the text). Measurements along the solid line
are shown in Figs.~1(g) and 3(a). (b) The ZBW calculation of the critical
current $I_\mathrm{C}$ of the DQD using the same parameters. The charge occupation of the
dots is indicated in parentheses. Visual representations
of a Cooper pair transfer when the DQD has an (c) even and (d) odd charge occupation. The $\pm 1$ values indicate the spin permutation parity for each spin
species, which yields a supercurrent reversal for an odd charge  occupation of
the DQD (see the text).}\label{fig2}
\end{figure}

In Fig.~2(a), we plot $I_\mathrm{DQD}$ as a function of the plunger gate voltages
$V_\mathrm{L}$ and $V_\mathrm{R}$, resulting in the zero magnetic field charge-stability diagram of the DQD mapped by the supercurrent. Remarkably, our
phase-sensitive measurement directly shows that the supercurrent reversal is
associated with the change in the total charge number, and it is absent in the
case of internal charge transfers with
$(n_\mathrm{L},n_\mathrm{R})\rightarrow(n_\mathrm{L}\pm1, n_\mathrm{R}\mp1)$.
However, $|I_\mathrm{DQD}|$ exhibits maxima near all charge boundaries,
consistently with earlier experiments \cite{PhysRevLett.121.257701}.

We understand these data using a two-orbital Anderson model, where each dot with
an on-site charging energy $U_i$ hosts a single spinful level at $\varepsilon_i$
with the dot index $i={\mathrm{L},\mathrm{R}}$. In the experiment, this
corresponds to a quantum dot orbital level spacing which is larger than the
charging energy \cite{Dam2006}. We consider an interdot charging energy term
$U_\mathrm{C} n_\mathrm{L} n_\mathrm{R}$ and an effective interdot tunneling
amplitude $t_\mathrm{C}$.
The tunnel coupling energies to the superconducting leads are denoted by
$\Gamma_\mathrm{L,R}$.

We consider the leading term of the supercurrent in the weak coupling limit
where $t_\mathrm{C}, \Gamma_\mathrm{L}, \Gamma_\mathrm{R} \ll \Delta \ll U_i$
\cite{DeFranceschi2010, doi:10.1080/00018732.2011.624266}, and evaluate the
current operator $I(\varphi)=i\frac{e}{\hbar}[H,n_\mathrm{R}]$, where $H$ is the
Hamiltonian of the system at a phase difference of $\varphi$ between the
superconducting leads (see the Supplementary Information \cite{supplement}). We numerically
evaluate $\langle I(\varphi)\rangle =I_\mathrm{C} \sin{\varphi}$ to find the
signful $I_\mathrm{C}$. We perform a global fit of the calculated sign reversal
contours [see the dashed lines in Fig.~2(a)] against the experimental dataset and
recover $U_\mathrm{L}=596.6\,\mu$eV, $U_\mathrm{R}=465.9\,\mu$eV,
$U_\mathrm{C}=41.5\,\mu$eV and $t_\mathrm{C}=85\,\mu$eV. We match the critical
current amplitude scale with the experimental data by setting
$\Gamma_\mathrm{L}=\Gamma_\mathrm{R}=33.2\,\mu$eV. The width of the even-odd
transitions establishes an upper bound on the electron temperature of the DQD,
$T<80\,$mK. We use these parameters to display
$I_\mathrm{C}(V_\mathrm{L},V_\mathrm{R})$ in Fig.~2b and find a good
correspondence with the experimental data using a zero bandwidth (ZBW)
approximation \cite{PhysRevB.68.035105,PhysRevLett.121.257701} (see the
Supplementary Information \cite{supplement}).

The observed supercurrent reversal \cite{spivak1991negative, Dam2006} is linked
to the number of permutations of fermion operators required to transfer a
spin-singlet Cooper pair through the DQD (see the Supplementary Information \cite{supplement}). In
the weak-coupling limit, this amounts to counting the number of same-spin dot
electrons, which each electron in the Cooper pair crosses. Each such crossing 
contributes with a factor of $-1$ to $I_\mathrm{C}$, which we illustrate for a
DQD with even [Fig.~2(c)], and odd charge occupations [Fig.~2(d)]. Consequently, the
sign of $I_\mathrm{C}$ is determined by the ground-state charge parity of the
DQD.

\begin{figure*}
\includegraphics[width=\linewidth]{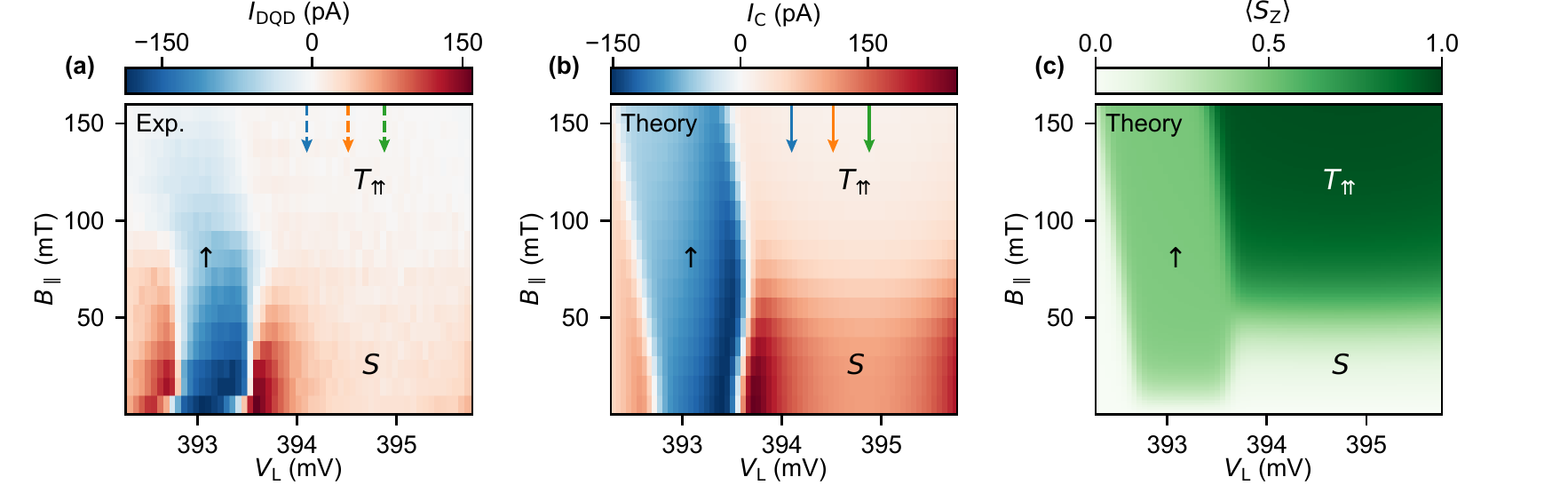}
\caption{The superconducting DQD in finite magnetic fields. (a) The measured
signful supercurrent oscillation amplitude $I_\mathrm{DQD}$ as a function of the
total energy [see the solid line in Fig.~2(a)] and magnetic field. Note the slight
charge shift between the zero magnetic field line and the rest of the data. (b)
The corresponding ZBW calculation of the signful critical current (see the text).
(c) The calculated spin expectation value in the ground state showing the
singlet to triplet transition in the even occupied state as a function of the
magnetic field. In (b) and (c), we use the parameters extracted in
Fig.~2b.}\label{fig3}
\end{figure*}

Next, we focus on the magnetic-field dependence of $I_\mathrm{DQD}$ [Fig.~3(a)]
along the total energy axis [solid line in Fig.~2(a)] spanning both even and odd
charge states. At $B_\parallel=0$, a finite  $t_\mathrm{C}$ results in a
singlet-triplet splitting $\Delta_\mathrm{ST}$ in the even occupied $(1,1)$
charge state \cite{RevModPhys.79.1217}. We model the DQD with an effective
identical $g$-factor on both dots, which results in a spin-polarized triplet
ground state above a threshold magnetic field,
$B_\mathrm{ST}=\Delta_\mathrm{ST}/(g^* \mu_\mathrm{B})$. To account for
spin-orbit coupling, we refine our interdot tunneling Hamiltonian to include
both spin-conserving and spin-flip tunneling amplitudes, $t_0$ and
$t_\textrm{x}$, resulting in an effective
$t_\mathrm{C}=\sqrt{t_0^2+t_\textrm{x}^2}$ (see the Supplementary Information \cite{supplement}).

With a global fit to the experimental data [Figs.~3(a) and (b)], we extract
$t_0=80\,\mu$eV, $t_\textrm{x}=30\,\mu$eV and $g^*=15.9$. This g-factor is in
agreement with earlier experimental values measured on InAs quantum dots
\cite{fasth2007direct, csonka2008giant, NadjPerge2010, PhysRevLett.107.176811}
and ballistic channels with superconducting leads \cite{van2017microwave,
PhysRevX.9.011010}. We estimate the spin-orbit length $l_\mathrm{SO} =
l_\mathrm{dot}t_0/(\sqrt{2}t_\textrm{x}) \approx 75\,$nm \cite{PhysRevB.88.075306}, using
the gate pitch as an estimate of the dot length, $l_\textrm{dot}=40\,$nm. This
coupling length yields an energy scale
$E_\textrm{SO}=\hbar^2/(2m^*l_\mathrm{SO}^2)=290\,\mu$eV with an effective
electron mass of $m^*=0.023m_e$, which is similar to earlier experimental
results on semiconductor nanowires in the presence of strong electrostatic
confinement \cite{PhysRevLett.108.166801, PhysRevB.94.035444}.

In Fig.~3(c), we plot the calculated expectation value $\langle S_Z \rangle$ of
the total spin $z$ component of the DQD, which visualizes the transition between
the spin singlet state $\langle S_Z \rangle=0$ and the spin-polarized triplet
state, where $\langle S_Z \rangle=1$, as a function of the magnetic field. This
transition point  at $B_\textrm{ST}$ is accompanied by a drop of the critical
current in the (1,1) sector, however this sudden decrease is absent in the odd
sector [see blue regions in Fig.~3(b)]. We note that the gradual global decrease
in $I_\mathrm{DQD}$ is consistent with the orbital effect of the magnetic field
applied along the nanowire \cite{PhysRevLett.119.187704}.

\begin{figure}
\includegraphics[width=\linewidth]{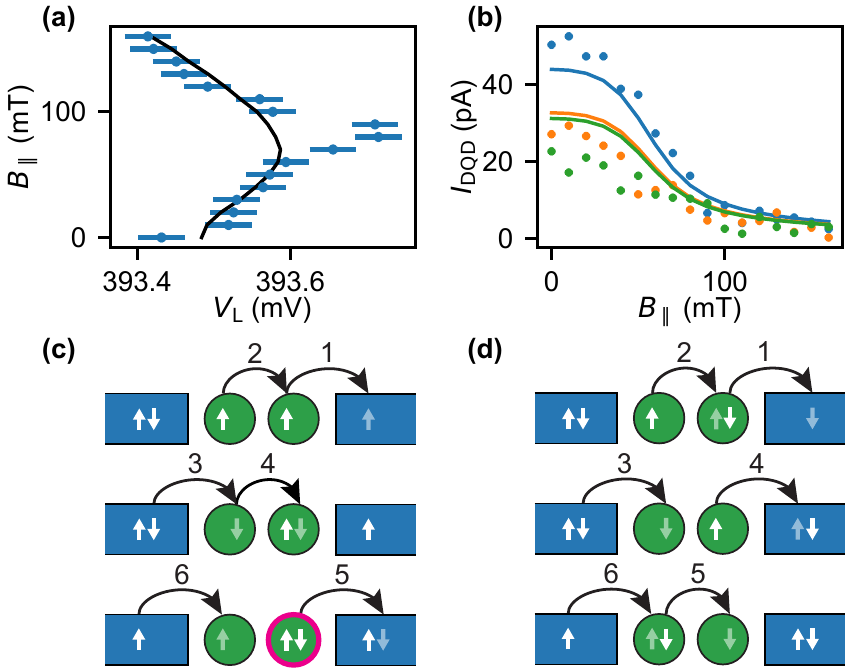}
\caption{Triplet-blockaded supercurrent. (a) The measured (blue dots and error
bars) and calculated (black solid line) even-odd charge boundary extracted from
Figs.~3(a) and 3(b). (b) Dots: The measured $I_\mathrm{DQD}$ at three plunger
gate settings in the even (1,1) sector [see the corresponding arrows in Fig.~3(a)].
Scaled theoretical values are shown as solid lines (see the text). Representative
sixth order tunneling processes are shown (c) in the $T_{\upuparrows}$ and (d) in
the singlet regime. The white arrows denote an initial occupied electron
state including the spin. The gray arrows visualize the final state for each
numbered process.}\label{fig4}
\end{figure}

We analyze these data in Fig.~4, where we first find the charge state boundary at
each value of $B_\parallel$ at $I_\mathrm{DQD}=0$ [blue dots and error bars in
Fig.~4(a)] and overlay the calculated boundary [black solid line, corresponding to
Fig.~3(b)]. We quantify $B_\mathrm{ST}\approx80\,$mT, which agrees consistently
with the characteristic cutoff magnetic field of $I_\mathrm{DQD}$ at several
plunger gate values [dots in Fig.~4b, colors corresponding to the arrows in
Fig.~3(a)]. However, we observe a deviation between the calculated and measured
charge boundary near $B_\mathrm{ST}$, which may stem from the microscopic details of
the spin-orbit coupling that our model does not account for. We
find an excellent agreement with the calculated critical current
$I_\mathrm{C}(B)$ [solid lines in Fig.~4(b)] with a common scaling factor of
$0.29$, which may be the result of the reduced switching current inside the charge
state due to thermal activation compared to the corresponding critical current
\cite{ivanchenko1969josephson}.

The suppression of the Josephson supercurrent through a DQD in the spin triplet
sector can be understood considering the virtual states involved in the Cooper
pair transfer. Starting from the $(1,1)$  $T_{\upuparrows}$ state close to the
charge boundary with the single occupation sector, we always encounter a virtual
state with a double occupation on one of the dots [magenta circle in Fig.~4(c)].
In the $U \gg \Delta$ limit corresponding to our experiments, this configuration
is energetically unfavorable and suppresses Cooper pair tunneling. In contrast,
a spin singlet starting condition can avoid this configuration [Fig.~4(d)]. We
finally note that the opposite limit, where $U \ll \Delta$, also leads to a
triplet supercurrent blockade \cite{droste2012josephson} (see the Supplementary
Information \cite{supplement}), which persists with a finite residual supercurrent in the spin
triplet state when $U \sim \Delta$.

In conclusion, we directly measured the supercurrent reversal associated with
the even-odd charge occupation in an InAs DQD, where the large level spacing
allows us to use a single orbital for each dot in our quantitative modeling. In
the $(1,1)$ charge sector, we showed that the singlet to triplet transition is
accompanied by a supercurrent blockade. This enables a direct spin to
supercurrent conversion \cite{PhysRevX.9.011010, hays2019continuous} in hybrid
semiconductor nanodevices \cite{DeFranceschi2010} used for quantum information
processing.

Raw datasets and computer code are available at the Zenodo repository
\cite{rawdata}.

\begin{acknowledgments}
The authors thank J. Danon for helpful discussions, A.~Proutski and D.~Laroche
for their technical input as well as J.~Mensingh, M.~Sarsby, O.~Benningshof and
R.~N.~Schouten for technical assistance. This work was supported by the QuantERA
project SuperTop, by the NANOCOHYBRI COST Action No.~CA16218, by the Netherlands
Organization for Scientific Research (NWO), by the Danish National Research
Foundation, by the National Research Development and Innovation Office of
Hungary within the Quantum Technology National Excellence Program (Project
No.~2017-1.2.1-NKP-2017-00001), under OTKA Grants No.~124723 and  No.~132146, and
the BME Nanotechnology and Materials Science TKP2020 IE grant (BME IE-NAT
TKP2020), and by the European Union's Horizon 2020 research and innovation
programme under grant No.~716655 (ERC Stg HEMs-DAM), No.~804988 (ERC Stg SiMS)
and No.~828948 (FET Open AndQC).
\end{acknowledgments}

\bibliography{main}

\end{document}